\documentclass[prd,twocolumn,nofootinbib]{revtex4}
\usepackage{graphicx, epsfig}
\usepackage{color}
\usepackage{mathrsfs,yhmath}
\usepackage{bm}
\usepackage{amsmath,amssymb}
\usepackage[stable]{footmisc}

\newcommand{\be}{\begin{equation}}
\newcommand{\ee}{\end{equation}}
\newcommand{\ba}{\begin{eqnarray}}
\newcommand{\ea}{\end{eqnarray}}

\newcommand{\gapp}{\mathrel{\raise.3ex\hbox{$>$}\mkern-14mu
              \lower0.6ex\hbox{$\sim$}}}
\newcommand{\lapp}{\mathrel{\raise.3ex\hbox{$<$}\mkern-14mu
              \lower0.6ex\hbox{$\sim$}}}

\newcommand{\bfk}{{\bf k}}

\newcommand{\bfp}{{\bf p}}

\newcommand{\bfx}{{\bf x}}

\definecolor{bittersweet}{rgb}{1.0, 0.44, 0.37}
\definecolor{coolblack}{rgb}{0.0, 0.18, 0.39}
\definecolor{britishracinggreen}{rgb}{0.0, 0.26, 0.15}
\definecolor{coolgrey}{rgb}{0.55, 0.57, 0.67}
\definecolor{darkgreen}{rgb}{0.0, 0.2, 0.13}
\definecolor{darkmagenta}{rgb}{0.55, 0.0, 0.55}
\definecolor{eggplant}{rgb}{0.38, 0.25, 0.32}
\definecolor{fashionfuchsia}{rgb}{0.96, 0.0, 0.63}

\begin{document}
\title{Unexciting classical backgrounds}
\author{Tanmay Vachaspati}
\affiliation{
Physics Department, Arizona State University, Tempe, AZ 85287, USA.
}

%%%%%%%%%%%%%%%%%%%%%%%%%%%%%%%%%%%%%%%%%%%%%%%%%%%%%%%
\begin{abstract}
\noindent
Quantum fields in time-dependent backgrounds generally lead to particle production. Here
we consider ``unexciting'' backgrounds for which the net particle production vanishes. 
We start by considering the simple harmonic oscillator and explicitly construct all unexciting 
time-dependent frequencies. This allows us to construct homogeneous backgrounds
in field theory
for which there is no particle production in any given mode, though we are able to show that
there are no homogeneous backgrounds for which the particle production vanishes in {\it every} 
mode. We then construct general inhomogeneous unexciting field theory backgrounds. The set of all 
unexciting field theory backgrounds will be further restricted by the choice of physical interactions 
and this  leads to an interesting open problem.
\end{abstract}
%%%%%%%%%%%%%%%%%%%%%%%%%%%%%%%%%%%%%%%%%%%%%%%%%%
%\pacs{???}

\maketitle

There has been considerable effort to study quantum radiation in time-dependent classical 
backgrounds ({\it e.g.}~\cite{Birrell:1982ix}). Landmark examples include Schwinger particle production~\cite{Schwinger:1951nm} 
and Hawking radiation~\cite{Hawking:1975vcx}. 
In the latter, the time dependence of the metric during gravitational collapse produces 
particles, while Schwinger particle production can be thought of as due to the time dependence of 
the electromagnetic vector gauge potential.

The present work is motivated by the Schwinger process for non-Abelian gauge fields
recently discussed in Ref.~\cite{Cardona:2021ovn} where a homogeneous non-Abelian 
electric field of a certain ``color''
produces (massless) gauge radiation of other colors. This process appears to be quite
general, so one might expect a similar process to occur even if the background electric
field is not uniform, for example if the color electric field is confined into flux tubes, as is
widely believed to occur in QCD. However, QCD electric flux tubes should not produce
quantum excitations if they are to be stable and confining. This motivates the general 
question -- can we find non-trivial space- and time-dependent backgrounds in which particle 
production does not occur?

\begin{figure}
\includegraphics[width=3.5in]{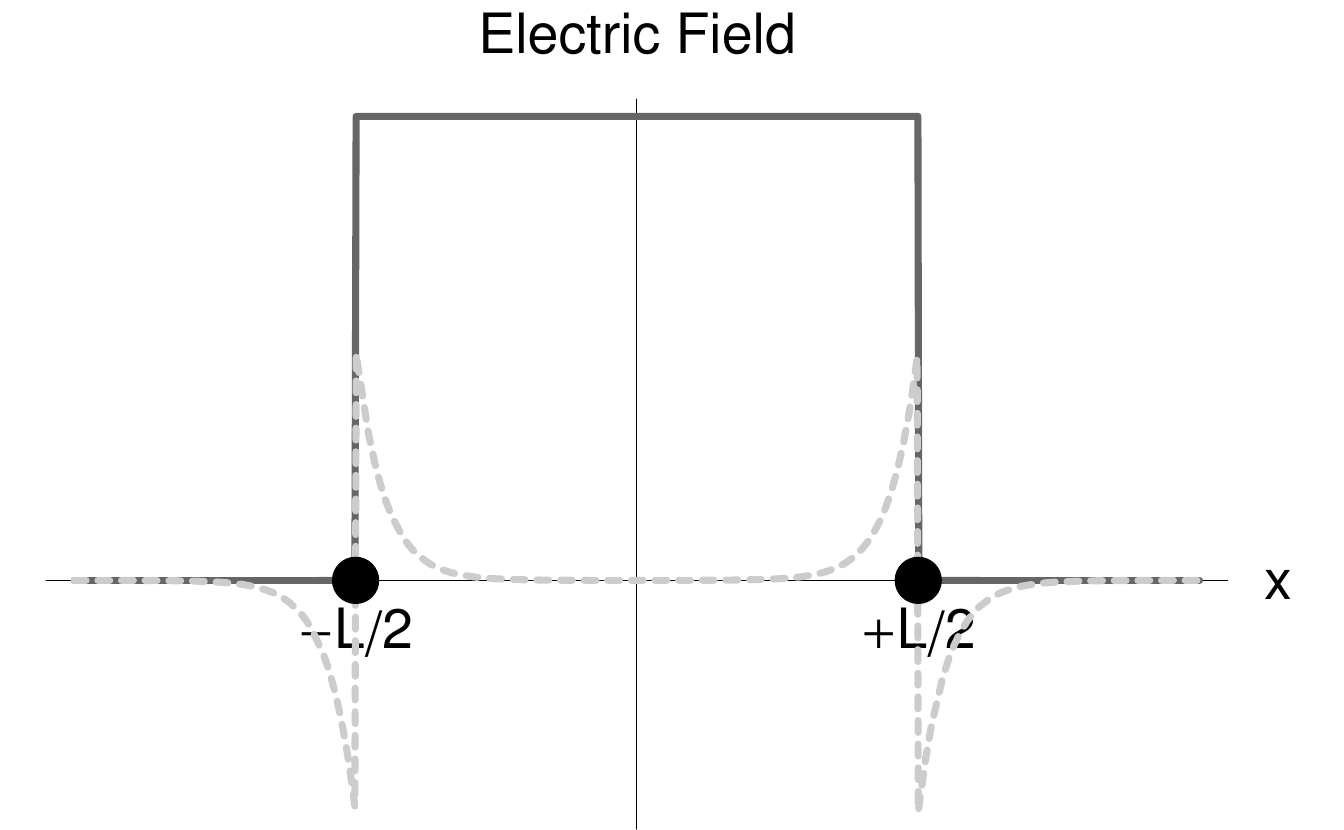}
\caption{As in Ref.~\cite{Chu:2010xc}, two infinitely heavy charges, $+Q$ and $-Q$ are 
placed at $x=-L/2$ and $x=+L/2$ respectively. The classical electric field is given by the 
thick dark line. With pair production, the electric field evolves into the
unexciting configuration illustrated by the dashed curve.}
\label{capacitor}
\end{figure}

An example of an unexciting electric field configuration is already known
in massless QED in 1+1 dimensions~\cite{Chu:2010xc,Gold:2020qzr}. 
One considers a capacitor consisting of external charges $+Q$ and $-Q$ separated 
by a distance $L$. The system can be solved completely since bosonization
yields a scalar plus gauge field theory with only bi-linear couplings.
The unexciting electric field background takes the form~\cite{Chu:2010xc},
\ba
F_{01} &=& Q ( \Theta (x+L/2) - \Theta (x-L/2) ) \nonumber \\
&& + g( f (x+L/2) - f(x-L/2) )
\label{efield}
\ea
where $g$ is the coupling constant in the model, 
and
\be
f(x) = -\frac{Q}{2g} {\rm sgn}(x) \left ( 1- e^{-g |x|} \right ).
\ee
A sketch of the unexciting electric field is shown as the dashed curve in Fig.~\ref{capacitor}.
Note, though, that the unexciting background is not purely an electric field as it also consists
of a condensate of fermion bound states. These bound states are described after bosonization
by a scalar field, $\phi$, that acquires a non-trivial profile,
\be
\phi(x) = f (x+L/2) - f(x-L/2).
\ee

Unexciting backgrounds may have practical utility as well. We can imagine situations
where a quantum system is in its ground state in a certain background ({\it e.g.} a magnetic 
field), and we would like to change the background to a final configuration while the
quantum system is finally in its ground state. The background would then have to be an
unexciting background.

The simple harmonic oscillator (SHO) with time-dependent frequency is the simplest system 
where this question can be analyzed (see also Ref.~\cite{sang2021}). 
Are there time-dependent frequencies for which the SHO does not get excited? 
In this case we are able to find a complete solution in Sec.~\ref{qsho}. To our surprise,
we find a very wide class of time-dependent frequencies, not necessarily adiabatic, for 
which particle production does not occur. 

The next step is to generalize the SHO result to quantum field theory in a classical background. 
We first consider homogeneous but time-dependent backgrounds. The homogeneity
of the background simplifies matters since the excitations can be diagonalized and each mode 
of the quantum field behaves as a quantum SHO with time-dependent frequency. In
Sec.~\ref{hombackgrounds}, we show that there are backgrounds for which particle production 
can be suppressed for at most a discrete set of modes and not for all modes. 

In Sec.~\ref{generalft} we consider the full problem of inhomogeneous, time-dependent
backgrounds. Here too we are able to find backgrounds for which there is no particle 
production. However, the solution does not address the constraint that only certain forms
of interactions may be present in a particular physical system. After discussing whether
field theory backgrounds may be unexciting at all times in Sec.~\ref{alltimes}, 
we turn to unexciting backgrounds that might arise in physical systems in 
Sec.~\ref{physical}. We are unable to construct a general unexciting {\it physical} background
in a field theory and leave it as an open problem.

\section{Quantum Simple Harmonic Oscillator}
\label{qsho}

Consider an SHO with unit mass $m=1$ and time
dependent frequency $\omega(t)$. We are interested in finding $\omega(t)$
such that there is no net energy production in quantum excitations.

Our analysis uses the ``classical-quantum correspondence'' (CQC) developed in
Refs.~\cite{Vachaspati:2018llo,Vachaspati:2018hcu} whereby quantum particle production 
in time-dependent backgrounds can be analyzed by solving a system of classical differential 
equations in higher dimensions. (The formalism only applies to bosonic particles.)
In the simplest case of a quantum SHO, the CQC maps the problem to a classical SHO in 
two dimensions, which can be described by a complex variable $z(t)$. Expectation values
of quantum operators can all be written as functions of $z$.

\subsection{SHO Solution}
\label{shosol}

The CQC equation for the complex variable $z(t)$ is
\be
z'' + \omega^2 z =0
\label{eom}
\ee
with initial conditions (taken at $t=t_i$)
\be
z_i = - \frac{i}{\sqrt{2\omega_i}}, \ \ 
z'_i = - \sqrt{\frac{\omega_i}{2}}
\label{ic}
\ee
where primes denote time derivatives and subscripts $i$ and $f$ refer to initial and 
final times. The energy in excitations is given by the function\footnote{There is no ambiguity 
in the definition of  excitation energy in contrast to 
the number of particles as discussed in the literature ({\it e.g.}~\cite{Dabrowski:2016tsx}
and, more recently,~\cite{Ilderton:2021zej}).}
\be
E(t) = \frac{1}{2} | z' - i\omega z |^2
\label{exen}
\ee
An unexciting background would be one for which the final energy
in excitations vanishes. 
Note that excitations may be produced and absorbed 
at intermediate times; we only require the final energy to vanish for the background
to be unexciting.
If instead, we require that the energy vanishes at all times, \eqref{exen} implies 
$z'=i\omega z$. Differentiating once and using \eqref{eom} implies $\omega'=0$. 
Hence there are no non-trivial backgrounds for which the SHO excitation energy 
vanishes for all times.

%\footnote{
%For the energy to vanish at all times, \eqref{exen} implies $z'=i\omega z$. Differentiating
%once and using \eqref{eom} implies $\omega'=0$. Hence there are no non-trivial backgrounds 
%for which the excitation energy vanishes for all times. \label{footnote1}}.

To derive an unexciting background we first write
\be
z(t) = \rho (t) e^{i\theta(t)}.
\ee
Then \eqref{eom} implies
\be
\rho'' + \omega^2 \rho = \frac{1}{4\rho^3}, \ \ 
\theta ' = - \frac{1}{2\rho^2}
\label{polareom}
\ee
where in the second equation we have used the initial conditions \eqref{ic} in terms
of $\rho$ and $\theta$,
\be
\rho_i = \frac{1}{\sqrt{2\omega_i}}, \ \ \rho'_i=0, \ \ 
\theta_i = \frac{3\pi}{2}, \ \ \theta '_i = -\omega_i.
\label{polaric}
\ee

Now we use the $\rho$ equation in \eqref{polareom} to solve for $\omega$ in
terms of $\rho$,
\be
\omega = \sqrt{\frac{1}{4\rho^4} - \frac{\rho ''}{\rho}}
\label{omsoln}
\ee
This tells us how the frequency should vary with time for any {\it choice} of $\rho (t) \ge 0$.
In addition, if we require $\omega^2 \ge 0$, then $4 \rho^3 \rho'' \le 1$, though
$\omega^2 < 0$ implies an inverted SHO potential and might be
acceptable for certain systems.

Now restrict the function $\rho (t)$ so that 
\be
\rho '_i = \rho ''_i =0, \ \ \rho '_f = \rho ''_f =0,
\label{shocond}
\ee 
while $\rho_i$ and $\rho_f$ are
unconstrained. For any choice of such $\rho (t)$, the energy in excitations
\be
E(t) = \frac{\rho'{}^2}{2}  + \frac{\rho^2}{2} \left ( \frac{1}{2\rho^2} - \omega \right )^2
\label{efunc}
\ee
satisfies
\be
E_i =0=E_f.
\ee
To see $E_i=0$, the initial conditions in \eqref{polaric} suffice.
To see $E_f=0$, note that the function $\rho$ is chosen to satisfy 
$\rho'_f=0$, so the first term in \eqref{efunc} vanishes, while
$\rho ''_f =0$ together with \eqref{omsoln} implies that the second
term in \eqref{efunc} vanishes at the final time.
Note that $\rho_i$ and $\rho_f$ can be different, which means
that $\omega_i$ and $\omega_f$ can be different. 

\begin{figure}
      \includegraphics[width=0.42\textwidth,angle=0]{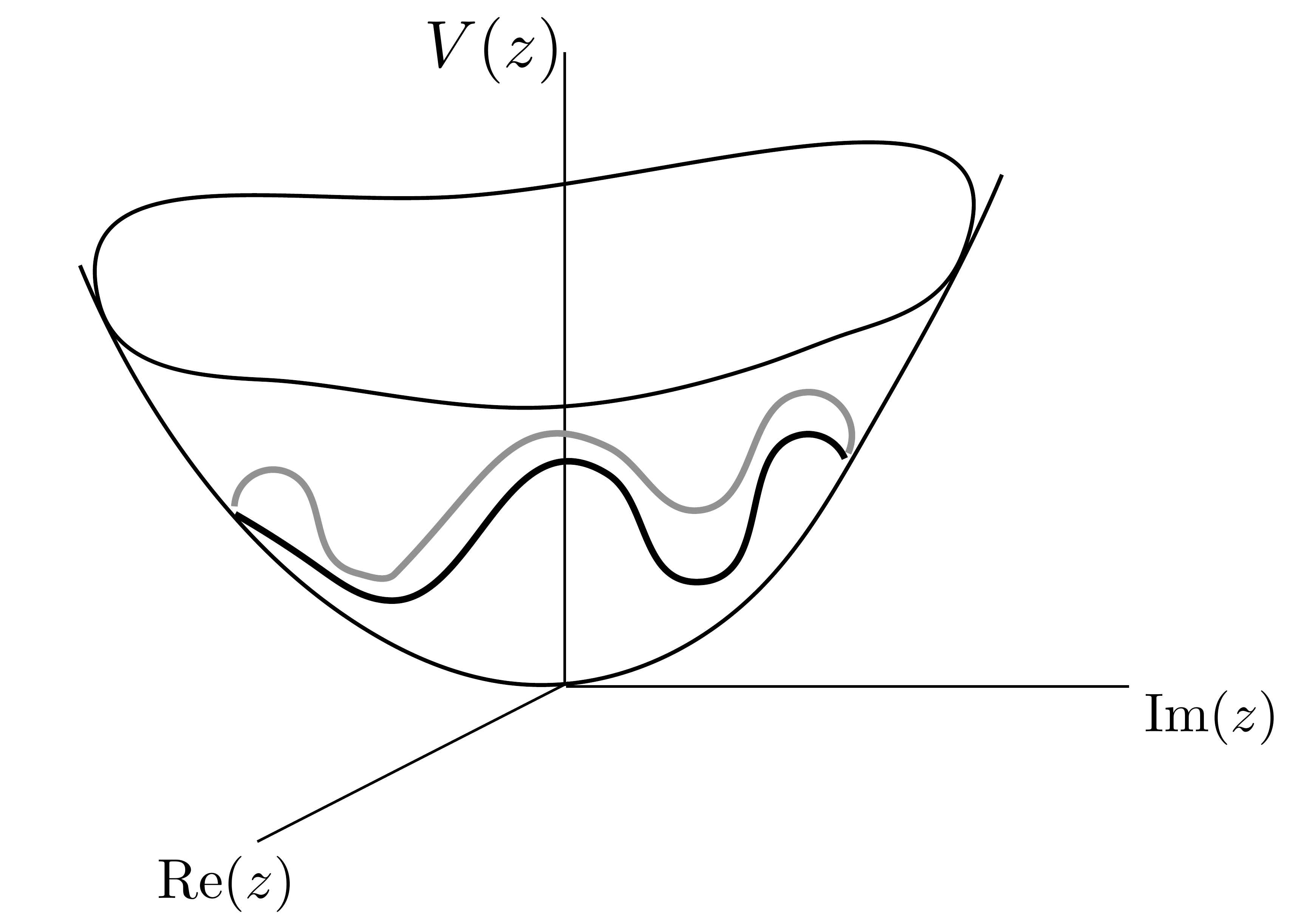}
        \caption{The potential for the complex variable $z$ is parabolic in two
        dimensions (one complex dimension). Conservation of angular momentum
        implies that the trajectory of $z$ goes around the parabola. If the trajectory
        oscillates, as shown by the solid curve, there is net particle production. 
        There is no net particle production if the trajectory does not oscillate at late times.}
  \label{rhosketch}
\end{figure}

In Fig.~\ref{rhosketch} we sketch the late time dynamics required for particle 
production.

Another quantity of interest may be the phase of the wavefunction, 
especially in cases where the final frequency equals the initial frequency.
The full wavefunction for the position $x$ of the simple harmonic oscillator 
can be written as
\be
\psi(t,x)= \frac{e^{i\gamma(t)}}{(2\pi\rho^2)^{1/4}} \exp \left [ \frac{i}{2}
\left ( \frac{\dot \rho}{\rho} + \frac{i}{2\rho^2} \right ) x^2 \right ]
\ee
where
\be
\gamma(t) = - \int_{t_i}^t \frac{dt'}{4\rho^2 (t')}.
\ee
Let us now consider the case of an unexciting background with 
$\omega_i = \omega_f$. Then the conditions in \eqref{shocond} imply
that $\rho_i=\rho_f$, and the phase difference from the case of a trivial
background with $\omega(t)=\omega_i$ is,
\be
\Delta \gamma(t) =  - \frac{1}{4}  \int_{t_i}^t dt' \left (  \frac{1}{\rho^2 (t')} - \frac{1}{\rho^2_i} \right ).
\ee

\subsection{An explicit example}
\label{explicit}

\begin{figure}
      \includegraphics[width=0.42\textwidth,angle=0]{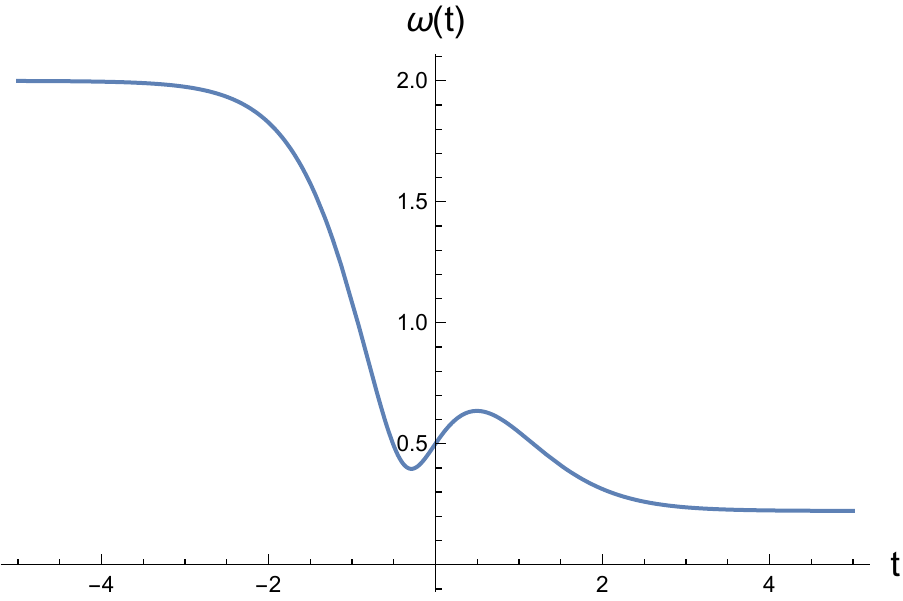}
        \caption{The frequency $\omega(t)$ for the explicit example 
        of Sec.~\ref{explicit}.}
  \label{omeplot}
\end{figure}

\begin{figure}
      \includegraphics[width=0.42\textwidth,angle=0]{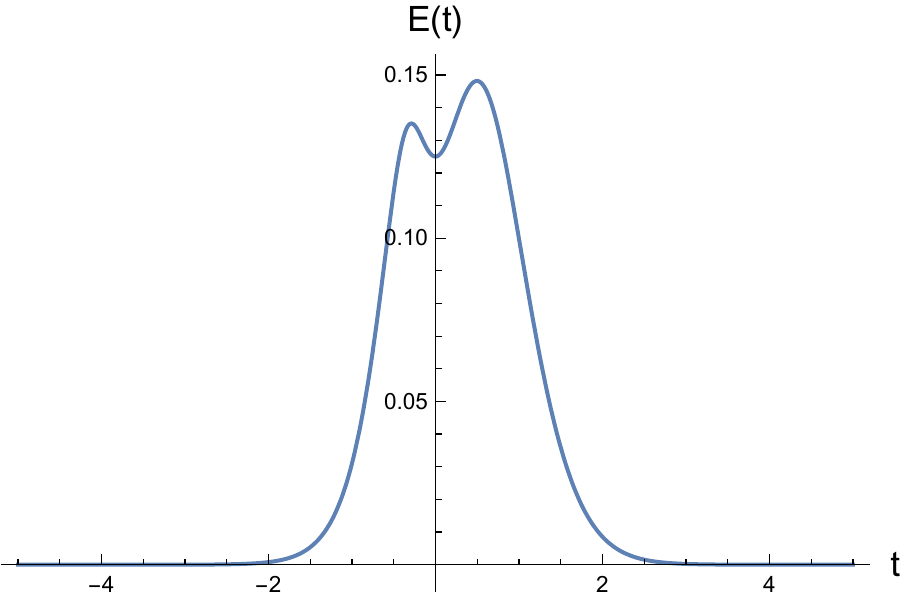}
        \caption{The excitation energy $E(t)$ for the explicit example 
        of Sec.~\ref{explicit}.}
  \label{eplot}
\end{figure}

Consider the choice of function
\be
\rho (t) = 1 + \frac{1}{2} \tanh(t).
\label{rhoexp}
\ee
with $t_i \to -\infty$ and $t_f \to +\infty$.
This choice satisfies the conditions 
$\rho'_i = \rho''_i =0$ and $\rho'_f = \rho''_f =0$
required for an unexciting background.
Then \eqref{omsoln} gives us $\omega (t)$ which we plot in Fig.~\ref{omeplot}
and in Fig.~\ref{eplot} we plot the energy in excitations as a function of time. 
At early times the energy in excitations grows but all the energy is absorbed 
at late times to give no net production of energy.

\subsection{A more general derivation}
\label{generalderiv}

The construction of the unexciting background in Sec.~\ref{shosol} was explicit
but it used polar coordinates that do not generalize easily to the field theory case.
Here we construct $\omega(t)$ in terms of the complex variables $z(t)$ and the
procedure can be generalized to field theory as in Sec.~\ref{generalft}.

We start with the identities
\be
(zz^*)' = zz'{}^* + z' z^*, \ \ 
(zz^*)'' = 2(z'z^*{}'-\omega^2 z z^*)
\label{zids}
\ee
Therefore,
\be
F(t) \equiv (z'-i\omega z)(z'+i\omega z)^* 
= \frac{1}{2} (zz^*)'' - i \omega (zz^*)' .
\label{ft}
\ee

Now consider $z(t)$ such that
\be
(zz^*)''_i = 0 = (zz^*)'_i
\label{zticond}
\ee
and
\be
(zz^*)''_f = 0 = (zz^*)'_f .
\label{ztfcond}
\ee
Then \eqref{ft} shows that $F_i=0=F_f$, implying that one of the two factors
$(z'-i\omega z)$ or $(z'+i\omega z)$ must vanish at $t_i$ and $t_f$.
At $t_i$ the initial conditions tell us that
\be
(z'-i\omega z)_i =0.
\ee

At $t_f$ we use the angular momentum constraint,
\be
z'z^* -z{z^*}{}' = i
\label{constraint}
\ee
and $\omega > 0$ to show that 
\be
|{z'}+i\omega z|^2 = |{z'}|^2 + \omega^2 |z|^2 + \omega > 0
\ee
provided $\omega > 0$. Therefore $({z'} + i\omega z)_f \ne 0$.
Then the only possibility is that
\be
({z'}-i\omega z)_f = 0.
\ee
However the energy in excitations is given by \eqref{exen}
and hence $E_f=0$ if we have a $z(t)$ such that \eqref{ztfcond} is satisfied. 
With such a choice of $z(t)$ we find $\omega$ as,
\be
\omega(t) = +\sqrt{-\frac{1}{2} \left ( \frac{z''}{z} +  \frac{z^*{}''}{z^*} \right ) } 
\label{omzsoln}
\ee
where we have made sure that the expression under the radical is real and we have 
only chosen the positive square root. 
With a little algebra, and making use of \eqref{constraint}, we recover
\eqref{omsoln}.

If we also require $\omega$ to be real valued, we must impose the condition
\be
-\frac{1}{2} \left ( \frac{z''}{z} +  \frac{z^*{}''}{z^*} \right )  \ge 0 .
\ee

To summarize, an unexciting background can be found from 
\be
\omega(t) = + \sqrt{-\frac{1}{2} \left ( \frac{z''}{z} +  \frac{z^*{}''}{z^*} \right ) }
\label{omzsoln}
\ee
by choosing any complex function $z(t)$ that satisfies the conditions
\eqref{zticond} and \eqref{ztfcond} together with the initial condition in 
\eqref{ic} and the Wronskian condition in \eqref{constraint}.

The solution in \eqref{omzsoln} is equivalent to the solution in \eqref{omsoln}
when written in terms of $\rho$ and $\theta$ together with the constraint in
\eqref{constraint}.

\section{Homogeneous backgrounds}
\label{hombackgrounds}

If the background is time-dependent but spatially homogeneous, the
quantum field can be expanded in Fourier modes and the problem reduces
to an infinite number of simple harmonic oscillators labeled by the wavenumber
of that mode. The time-dependent frequency of each mode is denoted $\omega_{\bf k}(t)$ 
and depends on the background under consideration. The variables corresponding to
the $z$'s for the single harmonic oscillator of Sec.~\ref{qsho} now carry the mode
index and will be written as $z_\bfk$. They satisfy the equation
\be
z_\bfk '' + \omega^2_\bfk z_\bfk =0
\ee
with initial conditions in \eqref{ic}. The frequencies $\omega_\bfk$ may take different
forms depending on the interactions in question. We will illustrate the arguments
for the form when a classical background field, $\phi (t)$, interacts with a quantum
field, $\psi(\bfx, t)$, due to a $\lambda\phi^2\psi^2/2$ interaction. Then,
\be
\omega^2_\bfk = \bfk^2 + \lambda \phi^2 (t).
\ee

From Sec.~\ref{qsho} we can certainly find a background for which a given mode
is not excited. But we are interested in finding a background for which {\it none}
of the modes is excited. Let us choose a background for which the mode $\bfk = \bfk_*$
is unexcited and denote the mode by $*$ subscripts. Then the background is given by 
\be
\omega^2_* (t) = -\frac{1}{2} \left ( \frac{z_* ''}{z_*} +  \frac{z^*{}''_*}{z^*_*} \right ) 
\label{omzsoln}
\ee
and
\be
\lambda \phi^2(t) = - \bfk^2_* - \frac{\rho_* ''}{\rho_*} + \frac{1}{4\rho_*^4}
\ee
Then, for another mode, say $\bfk =\bfp$, we must have
\be
z_\bfp '' + [ (\bfp^2 - \bfk^2_* ) + \omega^2_* (t)  ] z_\bfp =0
\ee
and initial conditions for $z_\bfp$ are as in \eqref{ic}. For $z_\bfp$ to
be unexcited, we require that $z_\bfp ''(t_f) =0$. However, the initial
conditions fix the evolution of $z_\bfp$ and the condition $z_\bfp '' (t_f)=0$
is an extra boundary condition on the evolution. In general, it will only be
satisfied for at most a discrete set of modes, not for all $\bfp$. Hence we
conclude that unexciting homogeneous backgrounds do not exist.
(Time-dependent electric fields with no Schwinger pair production
in a particular mode are discussed in~\cite{Kim:2011jw}.)

The story would be different if each mode of the quantum field were to
interact with an independent background. Then one would be able to
separately choose unexciting backgrounds for each mode. This suggests
that perhaps inhomogeneous backgrounds, where different background
modes couple to different excitation modes, can be unexciting. We now
turn to this question.

\section{General space and time dependent backgrounds}
\label{generalft}

A free quantum field in a general space and time dependent background can be
treated within the framework of the CQC. Then space is discretized, say with $N$
lattice points, and the Bogolyubov coefficients (generalized to inhomogeneous
backgrounds) correspond to an $N\times N$ matrix
that we denote by $Z$. The equation of motion for $Z$ is
\be
Z'' + \Omega^2 Z = 0
\label{Zeom}
\ee
where $\Omega =\Omega^\dag = \Omega^*$ contains both the spatial derivatives of 
the (real) field and the spacetime background.

The initial conditions for $Z$ are\footnote{To take the positive square root, the matrix under 
the radical is diagonalized, then the positive square root of each of the diagonal entries is taken, 
and finally the matrix diagonalization is inverted.},
\be
Z_i = -\frac{i}{\sqrt{2}} \bigr (\sqrt{\Omega_i} \, \bigr )^{-1} , \ \ 
Z'_i = \frac{1}{\sqrt{2}} \sqrt{\Omega_i}
\label{Zic}
\ee
where it is assumed that $\Omega_i$ is invertible.

The matrix $\Omega^2$ is a combination of the gradient terms for the quantum field
and its interactions with the background. Hence we write
\be
\Omega^2 = -\nabla^2 + U
\ee
where
\be
\nabla^2 = \begin{cases} -2/a^2, & i=j \\ 1/a^2, & i=j\pm 1 \\ 0, &{\rm otherwise}\end{cases}
\label{nabla2}
\ee
where $a$ is the lattice spacing. The form of the matrix $U$ is constrained by the form
of the interactions. For example, if the interactions are local, {\it i.e.} occur at the same
spatial point, then $U$ will be diagonal. For derivative interactions, $U$ will contain 
off-diagonal terms.

Further we have constraints that are satisfied by the evolution~\cite{Vachaspati:2018hcu},
\ba
Z^*{}' Z^T{}' - Z'Z^\dag {}' &=&0\label{C1} \\
Z^* Z^T - Z Z^\dag &=& 0\label{C2} \\
Z^* Z^T{}' - Z Z^\dag {}' &=& i. \label{C3}
\ea
These constraints can also be recast as~\cite{Vachaspati:2018hcu},
\ba
Z^\dag Z{}' - Z^\dag {}' Z &=& i \label{C4} \\
Z^\dag Z^* {}' - Z^\dag {}' Z^* &=& 0 . \label{C5}
\ea

The total energy in quantum excitations is given by
\be
E = \frac{1}{2} {\rm Tr} | Z' -i\Omega Z |^2
\label{Efield}
\ee
and we define an unexciting background to be one that gives $E_f =0$.

First we derive a necessary condition for an unexciting background. From
\eqref{Efield}, $E(t_f)=0$ implies
\be
Z_f' =i\Omega_f Z_f, \ \ Z_f^\dag {}' = -i Z_f^\dag \Omega_f .
\ee
Multiplying these two equations and simplifying gives
\be
(Z^\dag Z)''_f =0 .
\label{zzdagddot}
\ee
Eq.~\eqref{zzdagddot} is a
necessary condition to construct an unexciting background. Once we find a
suitable $Z$, the unexciting background is given by
\be
\Omega^2 = - \frac{1}{2} \left ( Z '' Z^{-1} + (Z^\dag)^{-1} Z^\dag {}''  \right )
\label{matdefomega}
\ee
and
\be
\Omega = + \sqrt{ - \frac{1}{2} \left ( Z '' Z^{-1} + (Z^\dag)^{-1} Z^\dag {}''  \right ) }
\label{Omegapositive}
\ee
where the $+$ sign indicates that the positive (matrix) root should be 
taken.

The condition \eqref{zzdagddot}, together with the constraints in \eqref{C1}, 
\eqref{C2} and \eqref{C3}, and the additional condition
\be
(Z^\dag Z)'_f =0
\label{zzdagdot}
\ee
are also {\it sufficient} for an unexciting background.
To show this, we rewrite \eqref{matdefomega} as
\be
\Omega^2 = - \frac{1}{2} (Z^\dag)^{-1} \left ( (Z^\dag Z)''  - 2 Z^\dag {}' Z' \right ) Z^{-1} 
\label{matdefomega2}
\ee
Having chosen some $Z(t)$, \eqref{matdefomega2} fixes $\Omega^2$ for all times.

We now show that \eqref{Omegapositive} gives vanishing energy at the final time.
This is because the condition in \eqref{zzdagddot} when inserted in \eqref{matdefomega2}
gives
\be
\Omega_f = + \sqrt{ (-i Z_f' Z^{-1}_f)^\dag (-i Z_f'  Z^{-1}_f) } 
\label{Omegaf1}
\ee

 Next we show that $M \equiv -i Z_f' Z^{-1}_f$ is Hermitian.
\be
M-M^\dag = -i (Z_f^\dag )^{-1} (Z^\dag Z )_f ' Z_f^{-1} =0,
\ee
since $Z(t)$ is chosen to satisfy \eqref{zzdagdot}. Therefore $M=M^\dag$. 
Further, using the constraint in \eqref{C4},
\be
M= \frac{1}{2} ( M+M^\dag ) = (Z Z^\dag)_f^{-1} 
\ee
and this is a positive matrix. Therefore \eqref{Omegaf1} gives
 \be
 \Omega_f =  -i Z_f' Z^{-1}_f = (Z Z^\dag)_f^{-1} 
 \label{Omegaf}
 \ee
 and so from \eqref{Efield},
 \be
E_f = \frac{1}{2} {\rm Tr} | Z_f' -i\Omega_f Z_f |^2 =0.
\label{Ef}
\ee
 
This proves
that to construct an unexciting background we can use \eqref{Omegapositive} where
$Z(t)$ satisfies the constraints in \eqref{C1}, \eqref{C2} and \eqref{C3}, and the final
time conditions in \eqref{zzdagddot} and \eqref{zzdagdot}.

\section{Solving the constraints}
\label{solvingconstraints}

Let us define
\be
\rho^2 = Z Z^\dag
\ee
where $\rho^2$ is real, symmetric and positive due to the constraint condition in \eqref{C2}.
Then we can write
\be
Z = \rho U
\label{ZrhoU}
\ee
where $U$ is a unitary matrix.
 
Now we turn to the constraint in \eqref{C3}. Insertion of \eqref{ZrhoU} in \eqref{C3} gives
the conditions
\be
[\rho, \rho ']=0,
\label{rhorhodot} 
\ee
\be
\{ \rho^2, U' U^\dag\} = i .
\label{rho2UUdag}
\ee
where the curly braces denote an anti-commutator.
Note that \eqref{rhorhodot} also implies $[\rho, \rho '']=0$.
 
A solution of Eq.~\eqref{rho2UUdag} is,
\be
U' U^\dag = \frac{i}{2} \rho^{-2}
\label{Ueq}
\ee
which is analogous to the solution in the case of a single simple harmonic oscillator
(see Eq.~\eqref{polareom})\footnote{The solution in \eqref{Ueq} is not unique. For
example, one could add any matrix on the right-hand side of \eqref{Ueq} that
anti-commutes with $\rho^2$.}.

With some algebra, we can check that $Z$ as given by \eqref{ZrhoU} satsfies
all the three constraints \eqref{C1}, \eqref{C2} and \eqref{C3}. With this $Z$ in
\eqref{matdefomega} we also
find
\be
\Omega^2 = - \rho '' \rho^{-1} + \frac{1}{4} \rho^{-4}
\label{Omega2rho}
\ee

So now the problem of constructing field theory unexciting backgrounds has been
reduced to suitably choosing a real-valued matrix function $\rho$ that satisfies
the conditions
\be
[\rho, \rho ' ]=0, \ \ \rho_i' =0= \rho_f ', \ \ \rho_i'' =0= \rho_f ''.
\label{rhoconditions}
\ee
Then we can construct $\Omega^2$ using \eqref{Omega2rho}.

A simple example solution is
\be
\rho (t) = A + \frac{1}{2} \tanh(t) B
\label{matrixrho}
\ee
where the time-independent, real, symmetric matrices $A$ and $B$ commute: 
$[A,B]=0$. This choice of $\rho$ satisfies all the conditions in \eqref{rhoconditions}
and from \eqref{Omega2rho} will lead to $\Omega^2$ that is unexciting.
The challenge however is to find $\rho(t)$ that not only gives an unexciting
background but is also consistent with interactions that are of physical interest.
We will turn to this question in Sec.~\ref{physical}.

\section{Unexciting for all times}
\label{alltimes}

In the case of the SHO it was simple to see that only the trivial background with
$\omega'=0$ is unexciting at all times as in Sec.~\ref{shosol}.
%(see Footnote~\ref{footnote1}).
Here we consider field theory backgrounds that may be unexciting for all times. 

Setting $E(t)=0$ in \eqref{Efield} gives
\be
Z' = i\Omega Z.
\label{ZpOmegaZ}
\ee
Differentiating once with respect to time and using \eqref{Zeom} gives,
\be
\Omega' Z =0.
\label{omegaz}
\ee
Assuming that $Z$ is invertible, this implies that $\Omega' =0$ and the time-dependence
of $\Omega$ is trivial.
An exception is when the background has some symmetries and there are excitation 
zero modes for then the initial $Z$ in \eqref{Zic} is not well-defined.

A second related exception is in situations where the background is ``stationary''. 
Then the background can have time-dependence but the spectrum of excitations is
time-independent and so $\Omega'=0$.
An example is when the background is due to a soliton, as discussed 
in~\cite{Mukhopadhyay:2021wmu}. 
A static soliton background has translational symmetry 
and the excitation spectrum has a zero mode~\cite{Vachaspati:2006zz}. 
A {\it boosted} soliton background is time-dependent but there is clearly no 
particle production since one can always go to the rest frame of the soliton.
%Then the initial 
%condition for $Z$ in \eqref{Zic} is singular as $\Omega_i$ has a zero eigenvalue and its 
%inverse is not defined. There is no quantum particle production if the soliton is simply 
%boosted even though the background is time-dependent.
A second similar example of an unexciting background is that of pp-waves~\cite{Deser:1975ffb}
and may be relevant for cosmologies with null big bang singularities~\cite{Craps:2005wd}.
A third example occurs in spontaneously broken non-Abelian gauge theories that
contain monopole solutions. Excitations of the rotor degree of freedom of the
monopole endows the monopole with electric charge and converts it into a 
dyon~\cite{Coleman:1982cx}.
The dyon fields are time-dependent but stationary, and the spectrum of excitations 
around a dyon is time-independent. This then brings us to the example of a pure
non-Abelian gauge theory. Here too there are rotor degrees of freedom whose time-dependence
produce stationary backgrounds~\cite{Brown:1979bv}. We plan to describe and analyze this 
example in a forthcoming publication.

Another important point to note is in the context of massless QED in 1+1 dimensions
mentioned in the introduction. There we have described an unexciting electric field
configuration. This background is unexciting {\it for all times} as no fermions are produced,
in contrast to our conclusion above. The reason is that our analysis using
the CQC only applies to the production of bosons and cannot be applied to fermionic
systems. Once the model is bosonized, the scalar field, $\phi$, couples directly to the 
electromagnetic field strength due to a $\phi \epsilon^{\mu\nu}F_{\mu\nu}$ coupling.
Even though the gauge potential is time dependent in temporal gauge, there can be 
no production of $\phi$ quanta in a static electric field background.

\section{Unexciting physical backgrounds?}
\label{physical}

The system of interest may be a quantum field interacting with a scalar background,
for example a $\lambda\phi^2\psi^2/2$ interaction as in Sec.~\ref{hombackgrounds}.
Or it could be charged particles interacting with a background electric field, as in
Schwinger particle production. Or it could be both a scalar field and an electric field,
and also perhaps a gravitational background. Depending on the system, the form of
the frequency matrix $\Omega^2$ is restricted and it is of interest to find unexciting
backgrounds consistent with the interactions of interest.

Let us illustrate the problem with our example from Sec.~\ref{hombackgrounds}
where the interaction is $\lambda\phi^2 \psi^2/2$ and $\phi (t,{\bf x})$ is the space 
and time dependent background. In this case the interaction acts like an effective 
mass term and $\Omega^2$ takes the form,
\be
\Omega^2 = - \nabla^2 + (m^2 + \lambda \phi^2 ) 
\label{Oform}
\ee
where $\nabla^2$ is given in \eqref{nabla2} and is a symmetric, tri-diagonal matrix, 
while the $m^2 + \lambda \phi^2$ term is a diagonal matrix. 
From \eqref{Omega2rho} we can write,
\be
 \lambda \phi^2 = - \left ( \square \rho + m^2\rho   - \frac{1}{4} \rho^{-3} \right ) \frac{1}{\rho} 
 \label{lphi2}
\ee
where $\square = \partial_t^2 - \nabla^2$ is the D'Alembertian (matrix) operator.
Since $\lambda\phi^2$ has to be a diagonal matrix, this imposes an additional constraint
on $\rho$, namely that the right-hand side of \eqref{lphi2} be diagonal. 
It is not clear how to choose a non-trivial $\rho(t)$ that satisfies \eqref{rhoconditions} 
and that leads to a diagonal form for $\lambda\phi^2$ in \eqref{lphi2}. 

The physical system of quantum excitations in a color electric field~\cite{Cardona:2021ovn}
is similar to that of the scalar field discussed above but with additional complications due
to group indices and three spatial dimensions.
The background vector gauge potential can be taken in temporal gauge to be 
$A^a_i = E_i ({\bf x}) f(t) \delta^{a3}$, 
where $E_i ({\bf x})$ is the chosen background electric field function of the $a=3$ color,
and the function $f(t)$ is chosen to suitably turn the electric field on and off asymptotically. 
(We assume that external currents are present so that the background magnetic field 
vanishes.) The leading interaction between the background and the gluonic excitations
will again be local; only the gradient terms provide couplings of the excitation fields
at different spatial points. The analog of \eqref{lphi2} for this problem will again
require that a matrix $\rho$ be chosen so that a combination similar to that on the 
right-hand side of \eqref{lphi2} be diagonal.

\section{Conclusions}
\label{conclusions}

Our investigations were motivated by Schwinger pair production in the
background of a non-Abelian electric field, but the question is more general -- 
are there classical time-dependent backgrounds that do not produce quantum excitations?

To address this question,
we first considered a quantum simple harmonic oscillator with a time-dependent 
frequency. We found an infinite set of unexciting backgrounds --
variations of the frequency, even possibly rapid, that lead to no net production of
excitations. The result is potentially of interest in practical settings where one may
wish to alter external backgrounds without disturbing a quantum system.

We then considered the quantum field theory case. The spatially homogeneous
background problem can be diagonalized and becomes equivalent to an infinite set of simple 
harmonic oscillators. We argued that we could suppress excitations of some modes
by choosing a suitable background time-dependence. However, there are always some
modes that get excited by the time-dependent background and hence a homogeneous 
background cannot be unexciting.

Finally we considered the general case of inhomogeneous, time-dependent backgrounds.
Here we were able to derive a formula that enables us to construct unexciting backgrounds. 
However these are ``idealized'' backgrounds and, as discussed in Sec.~\ref{physical}, may 
not correspond to physical interactions, {\it e.g.} an electric field background. The question
whether there are unexciting {\it physical} backgrounds is still open, one we hope to
return to in the future. 

Another question of interest that we considered in Sec.~\ref{alltimes} is if there are 
classical time-dependent backgrounds that are unexciting for all times. We showed that 
such backgrounds may exist in bosonic systems provided the background has symmetries
and the time-dependence is purely in the variables that are conjugate to the symmetry 
generators. For then, the time-dependence leads to a stationary background in which 
the spectrum of excitations is time-independent and hence there is no particle production.
A related question is to find backgrounds in which excitations 
are continuously created and absorbed in an oscillatory fashion, with no net production 
on average.

\acknowledgements
I thank George Zahariade for many helpful discussions and Carlos Cardona, Sang Pyo Kim,
Mainak Mukhopadhyay, Christian Schubert and Savdeep Sethi for comments.
This work was supported by the U.S. Department of Energy, Office of High Energy Physics, 
under Award DE-SC0019470 at ASU.

\bibstyle{aps}
\bibliography{paper}

\end{document}